\documentclass[12pt]{article}
\usepackage{graphicx}
\usepackage{epsfig}
\usepackage{epstopdf}
\DeclareGraphicsExtensions{.pdf,.eps,.png,.jpg,.mps}

\usepackage{cite}

\def\lsim{\mathrel{\rlap{\lower3pt\hbox{\hskip0pt$\sim$}}
     \raise1pt\hbox{$<$}}}         
\def\gsim{\mathrel{\rlap{\lower4pt\hbox{\hskip1pt$\sim$}}
     \raise1pt\hbox{$>$}}}         

\usepackage{amsmath}
\usepackage{amsfonts}

\begin{document}
\begin{titlepage}

\centerline{\Large \bf How to Combine a Billion Alphas}
\medskip

\centerline{Zura Kakushadze$^\S$$^\dag$\footnote{\, Zura Kakushadze, Ph.D., is the President of Quantigic$^\circledR$ Solutions LLC,
and a Full Professor at Free University of Tbilisi. Email: zura@quantigic.com} and Willie Yu$^\sharp$\footnote{\, Willie Yu, Ph.D., is a Research Fellow at Duke-NUS Medical School. Email: willie.yu@duke-nus.edu.sg}}
\bigskip

\centerline{\em $^\S$ Quantigic$^\circledR$ Solutions LLC}
\centerline{\em 1127 High Ridge Road \#135, Stamford, CT 06905\,\,\footnote{\, DISCLAIMER: This address is used by the corresponding author for no
purpose other than to indicate his professional affiliation as is customary in
publications. In particular, the contents of this paper
are not intended as an investment, legal, tax or any other such advice,
and in no way represent views of Quantigic$^\circledR$ Solutions LLC,
the website \underline{www.quantigic.com} or any of their other affiliates.
}}
\centerline{\em $^\dag$ Free University of Tbilisi, Business School \& School of Physics}
\centerline{\em 240, David Agmashenebeli Alley, Tbilisi, 0159, Georgia}
\centerline{\em $^\sharp$ Centre for Computational Biology, Duke-NUS Medical School}
\centerline{\em 8 College Road, Singapore 169857}
\medskip
\centerline{(February 27, 2016)}

\bigskip
\medskip

\begin{abstract}
{}We give an explicit algorithm and source code for computing optimal weights for combining a large number $N$ of alphas. This algorithm does not cost ${\cal O}(N^3)$ or even ${\cal O}(N^2)$ operations but is much cheaper, in fact, the number of required operations scales linearly with $N$. We discuss how in the absence of binary or quasi-binary ``clustering" of alphas, which is not observed in practice, the optimization problem simplifies when $N$ is large. Our algorithm does not require computing principal components or inverting large matrices, nor does it require iterations. The number of risk factors it employs, which typically is limited by the number of historical observations, can be sizably enlarged via using position data for the underlying tradables.
\end{abstract}
\medskip
\end{titlepage}

\newpage
\section{Introduction and Summary}

{}Now that machines have taken over alpha\footnote{\, Here ``alpha" -- following the common trader lingo -- generally means any reasonable ``expected return" that one may wish to trade on and is not necessarily the same as the ``academic" alpha.  In practice, often the detailed information about how alphas are constructed may not even be available, e.g., the only data available could be the position data, so ``alpha" then is a set of instructions to achieve certain stock (or some other instrument) holdings by some times $t_1,t_2,\dots$} mining, the number of available alphas is growing exponentially. On the flip side, these ``modern" alphas are ever fainter and more ephemeral. To mitigate this effect, among other things, one combines a large number of alphas and trades the so-combined ``mega-alpha". And this is nontrivial.

{}Why? It is important to pick the alpha weights optimally, i.e., to optimize the return, Sharpe ratio and/or other performance characteristics of this alpha portfolio. The commonly used techniques in optimizing alphas are conceptually similar to the mean-variance portfolio optimization \cite{Markowitz} or Sharpe ratio maximization \cite{Sharpe94} for stock portfolios. However, there are some evident differences.\footnote{\, In the olden days the alpha weights would have to be nonnegative. In many practical applications this is no longer the case as only the ``mega-alpha" is traded, not the individual alphas.} The most prosaic difference is that the number of alphas can be huge, in hundreds of thousands, millions or even billions. The available history (lookback), however, naturally is much shorter. This has implications for determining the alpha weights.

{}Let us look at vanilla Sharpe ratio maximization of the alpha portfolio with weights $w_i$, $i=1,\dots,N$, where $N$ is the number of alphas.\footnote{\, With no position bounds, trading costs, etc. -- these do not affect the point we make here.} The optimal weights are given by
\begin{equation}\label{weights}
 w_i = \eta \sum_{j=1}^N C^{-1}_{ij}~E_j
\end{equation}
where $E_j$ are the expected returns for our alphas, $C^{-1}_{ij}$ is the inverse of the alpha return covariance matrix $C_{ij}$, and $\eta$ is the normalization coefficient such that
\begin{equation}\label{norm}
 \sum_{i=1}^N \left|w_i\right| = 1
\end{equation}
If we compute $C_{ij}$ as a sample covariance matrix based on a time series of realized returns (see (\ref{sample.cov.mat})), it is badly singular as the number of observations is much smaller than $N$. This also happens in the case of stock portfolios. In that case one either builds a proprietary risk model to replace $C_{ij}$ or opts for a commercially available (multifactor) risk model. In the case of alphas the latter option is simply not there.

{}So, what is one to do? We can try to build a risk model for alphas following a rich experience with risk models for stocks. In the case of stocks a more popular approach is to combine style risk factors (i.e., those based on measured or estimated properties of stocks, such as size, volatility, value, etc.) and industry risk factors (i.e., those based on stocks' membership in sectors, industries, sub-industries, etc., depending on the nomenclature used by a particular industry classification employed). The number of style factors is limited, of order 10 for longer-horizon models, and about 4 for shorter-horizon models. In the case of stocks, at least for shorter-horizon models, it is the ubiquitous industry risk factors (numbering in a few hundred for a typical liquid trading universe) that add most value. However, there is no analog of the (binary or quasi-binary) industry classification for alphas. In practice, for many alphas it is not even known how they are constructed, only the (historical and desired) positions are known. Even formulaic alphas \cite{KLT} are mostly so convoluted that realistically it is impossible to classify them in any meaningful way, at least not such that the number of the resulting (binary or quasi-binary)\footnote{\, By binary ``clusters" we mean that each alpha would belong to one and only one ``cluster". By quasi-binary clusters we mean that this would be mostly the case but a (small) fraction of alphas could possibly belong to multiple (at most several) ``clusters". This is analogous to binary and quasi-binary (i.e., where we have some conglomerates belonging to multiple industries, sub-industries, etc., depending on the naming conventions) industry classifications in the case of stocks.} ``clusters" would be numerous enough to compete with principal components (see below).\footnote{\, There is also the issue of stability. Stocks rarely, if ever, jump industries rendering well-constructed industry classifications quite stable. However, alphas being ephemeral objects make poor candidates for being classified into any kind of stable binary or quasi-binary ``clusters".} And there are only a few a priori relevant style factors for alphas \cite{AlphaFM} to compete with the principal components.

{}Just as in the case of stocks, we can resort to the principal components of the sample covariance (or correlation) matrix to build our multifactor risk model for alphas. This is where one of our key observations comes in. As we discuss in detail below, irrespective of how a factor model for $C_{ij}$ is built, in the absence -- which we assume based on our discussion above -- of (binary or quasi-binary) ``clustering", when the number of alphas $N$ is large, the optimization (\ref{weights}) invariably reduces to a (weighted) regression! This also holds for any reasonably realistic deformation (e.g., shrinkage \cite{LW}) of the sample covariance matrix for such deformations can be viewed as multifactor risk models. I.e., there is no need to construct a full-blown risk model and compute the factor covariance matrix or even the specific (idiosyncratic) risk.\footnote{\, More precisely, we can do that, but in the 0th -- and very good -- approximation we need not.} So, as the simplest variant, we construct the factor loadings matrix from the first $M$ principal components of the sample covariance matrix corresponding to its positive eigenvalues and regress expected returns for alphas over this factor loadings matrix with the regression weights given by the inverse sample variances $C_{ii} = \sigma_i^2$. However, it turns out that we do not even need to calculate the principal components thereby further reducing computational cost.

{}Here is a simple prescription for obtaining the weights $w_i$. Start with a time series $R_{is}$ of realized alpha returns ($s=1,\dots,M+1$ labels the times $t_s$). Calculate the sample variances $C_{ii} = \sigma_i^2$ (but not the sample correlations). This costs ${\cal O}(MN)$ operations. Normalize the returns via ${\widetilde R}_{is} = R_{is}/\sigma_i$. Demean ${\widetilde R}_{is}$ both cross-sectionally and serially. Let us call the so-demeaned returns $Y_{is}$. Take only $M-1$ columns in $Y_{is}$ (e.g., the first $M-1$ columns). Take the expected returns $E_i$ for the alphas and normalize them via ${\widetilde E}_i = E_i/\sigma_i$. Regress ${\widetilde E}_i$ over the $N\times (M-1)$ matrix $Y_{is}$ with unit weights and no intercept. This regression costs ${\cal O}(M^2 N)$ operations. Take the residuals ${\widetilde \varepsilon}_i$ of this regression. Then the optimal weights $w_i = \eta{\widetilde \varepsilon}_i/\sigma_i$, where $\eta$ is fixed via (\ref{norm}). If the reader is only interested in the prescription and the source code, then the reader can go straight to Appendix \ref{app.A}, where we give R source code for this algorithm,\footnote{\, R Package for Statistical Computing, http://www.r-project.org. The source code given in Appendix \ref{app.A} is not written to be ``fancy" or optimized for speed or in any other way. Its sole purpose is to illustrate the algorithms described in the main text in a simple-to-understand fashion. Some legalese relating to this code is given in Appendix \ref{app.C}.} and skip the rest of the paper. However, if the reader would like to understand why this algorithm makes sense and also, among other things, how to potentially increase the number of risk factors from $M$ (for a generous 1 year lookback $M \approx 250$) to a few thousand, then the reader may wish to keep reading.

{}To summarize, calculating the weights $w_i$ does not require computing any principal components or inverting any large matrices, and it costs only ${\cal O}(M^2 N)$ operations, so it is linear in $N$. Furthermore, the algorithm is not iterative, so there are no convergence issues to worry about. Perhaps somewhat ironically, the simplicity of this algorithm is rooted in the very nature of this problem, that we have a small number of observations compared with the large (in fact, huge) number of alphas. As we discuss in more detail in the subsequent sections, there is simply not much else one can do that would be out-of-sample stable with the exception of enlarging the number of risk factors provided there is additional information available to us.

{}The remainder of this paper is organized as follows. In Section \ref{sec.2} we set up our framework and notations. In Section \ref{sec.3} we discuss why, absent ``clustering", optimization using a factor model reduces to a regression in the large $N$ limit. In Section \ref{sec.4} we discuss why this also applies to deformations of the sample covariance matrix as they too reduce to factor models. We also discuss why style factors add little value and how to enlarge the number of risk factors based on more detailed position data (as opposed to the historical alpha return data). In Section \ref{sec.5} we discuss a refinement whereby the analog of the ``market" mode for stocks is factored out to improve performance of the alpha portfolio. We also give the detailed algorithm for obtaining the optimal alpha weights and discuss the computational cost (including why it is cheaper than doing principal components). We briefly conclude in Section \ref{sec.6}. Appendix \ref{app.A} contains the source code for our algorithm. Appendix \ref{app.C} contains some legalese. Parts of Sections 3 and 4 are based on \cite{KYb}.

\section{Sample Covariance Matrix}\label{sec.2}

{}So, we have $N$ time series of returns. A priori these returns can correspond to stocks or some other instruments, alphas, etc. Here we will be general and refer to them simply as returns, albeit we will make some assumptions about these returns below. Each time series contains $M+1$ observations corresponding to times $t_s$, and we will denote our returns as $R_{is}$, where $i=1,\dots,N$ and $s=1,\dots,M,M+1$ ($t_1$ is the most recent observation). The sample covariance matrix (SCM) is given by\footnote{\, The overall normalization of $C_{ij}$ does not affect the weights $w_i$ in (\ref{weights}), so the difference between the unbiased estimate with $M$ in the denominator vs. the maximum likelihood estimate with $M+1$ in the denominator is immaterial for our purposes here. Also, in most applications $M \gg 1$.}
\begin{equation}\label{sample.cov.mat}
 C_{ij} = {1\over M}\sum_{s=1}^{M+1} X_{is}~X_{js}
\end{equation}
where $X_{is} = R_{is} - {\overline R}_i$ are serially demeaned returns; ${\overline R}_i = {1\over {M+1}}\sum_{s=1}^{M+1} R_{is}$.

{}We are interested in cases where $M < N$, in fact, $M\ll N$. When $M < N$, $C_{ij}$ is singular: we have $\sum_{s=1}^{M+1} X_{is} = 0$, so only $M$ columns of the matrix $X_{is}$ are linearly independent. Let us eliminate the last column: $X_{i,M+1}=-\sum_{s=1}^M X_{is}$. Then we can express $C_{ij}$ via the first $M$ columns:
\begin{equation}\label{SCM}
 C_{ij} = \sum_{s,s^\prime=1}^{M} X_{is}~\phi_{ss^\prime}~X_{js^\prime}
\end{equation}
Here $\phi_{ss^\prime} = \left(\delta_{ss^\prime} + u_s u_{s^\prime}\right)/M$ is a nonsingular $M\times M$ matrix ($s,s^\prime = 1,\dots,M$); $u_s \equiv 1$ is a unit $M$-vector. Note that $\phi_{ss^\prime}$ is a 1-factor model (see below).

{}The challenge is to either deform $C_{ij}$ such that it is nonsingular, or to replace it with a constructed nonsingular matrix $\Gamma_{ij}$ such that it reasonably approximates $C_{ij}$ in-sample and predicts it out-of-sample. Let us first discuss the latter approach.

\section{Factor Models}\label{sec.3}

{}A popular method -- at least in the case of equities -- for constructing a nonsingular replacement $\Gamma_{ij}$ for $C_{ij}$ is via a factor model:\footnote{\, For equity multifactor models, see, e.g., \cite{GK} and references therein. A multifactor model approach for alphas was set forth and discussed in detail in \cite{AlphaFM}.}
\begin{equation}\label{fac.mod}
 \Gamma_{ij} = \xi_i^2~\delta_{ij} + \sum_{A,B=1}^K \Omega_{iA}~\Phi_{AB}~\Omega_{jB}
\end{equation}
Here: $\xi_i$ is the specific (a.k.a. idiosyncratic) risk for each return; $\Omega_{iA}$ is an $N\times K$ factor loadings matrix; and $\Phi_{AB}$ is a $K\times K$ factor covariance matrix (FCM), $A,B=1,\dots,K$. The number of factors $K\ll N$ to have FCM more stable than SCM.

{}The nice thing about $\Gamma_{ij}$ is that it is positive-definite (and therefore invertible) if FCM is positive-definite and all $\xi_i^2 > 0$. For our purposes here it is convenient to rewrite $\Gamma_{ij}$ via $\Gamma_{ij} = \xi_i~\xi_j~\gamma_{ij}$, where
\begin{equation}
 \gamma_{ij} = \delta_{ij} + \sum_{A=1}^K \beta_{iA}~\beta_{jA}
\end{equation}
and $\beta_{iA} = {\widetilde \beta}_{iA} / \xi_i$. Here (in matrix notations) ${\widetilde\beta} = \Omega~{\widetilde \Phi}$, and ${\widetilde\Phi}$ is the Cholesky decomposition of $\Phi$, so ${\widetilde\Phi}~{\widetilde \Phi}^T = \Phi$. The weights (\ref{weights}) (with $C_{ij}$ replaced by $\Gamma_{ij}$) are given by
\begin{equation}
 w_i = {\eta\over\xi_i}\sum_{j = 1}^N \gamma_{ij}^{-1}~{E_j\over\xi_j} = {\eta\over\xi_i}\left[{E_i\over\xi_i} - \sum_{j = 1}^N \sum_{A,B = 1}^K \beta_{iA}~Q^{-1}_{AB}~\beta_{jB}~{E_j\over\xi_j}\right]
\end{equation}
where $Q_{AB} = \delta_{AB} + q_{AB}$, and $q_{AB} = \sum_{i=1}^N \beta_{iA}~\beta_{iB}$. The diagonal elements of this matrix are $Q_{AA} = 1 + \sum_{i=1}^N \beta_{iA}^2$. It then follows that, if all $q_{AA} = \sum_{i=1}^N \beta_{iA}^2 \gg 1$, which we will argue to be the case momentarily, then $Q_{AB}\approx q_{AB} = \sum_{i=1}^N z_i~{\widetilde\beta}_{iA}~{\widetilde\beta}_{iB}$ and
\begin{equation}\label{w.reg}
 w_i \approx \eta~z_i\left[E_i - \sum_{j = 1}^N \sum_{A,B = 1}^K {\widetilde \beta}_{iA}~q^{-1}_{AB}~{\widetilde \beta}_{jB}~z_j ~E_j\right] =
 \eta~z_i~\varepsilon_i
\end{equation}
where $\varepsilon_i$ are the residuals of the cross-sectional weighted regression\footnote{\, Without the intercept unless it is subsumed in a linear combination of the columns of ${\widetilde \beta}_{iA}$.} of $E_i$ over ${\widetilde \beta}_{iA}$ with the weights $z_i = 1/\xi_i^2$, or, equivalently, ${\widetilde \varepsilon}_i = \varepsilon_i/\xi_i$ are the residuals of the unit-weighted regression of ${\widetilde E}_i = E_i / \xi_i$ over $\beta_{iA}$. So, (\ref{weights}) reduces to a regression.

{}The question is why -- or, more precisely, when -- all $q_{AA}\gg 1$. This is the case when: i) $N$ is large, and ii) there is no ``clustering" in the vectors $\beta_{iA}$. That is, we do not have vanishing or small values of $\beta_{iA}^2$ for most values of the index $i$ with only a small subset thereof having $\beta_{iA}^2\gsim 1$. Without ``clustering", to have $q_{AA}\lsim 1$, we would have to have $\beta^2_{iA}\ll 1$, i.e., $\gamma_{ij}$ and consequently $\Gamma_{ij}$ would be almost diagonal.

{}E.g., consider a 1-factor model ($K=1$) with uniform $\beta_i \equiv \beta$. In this model we have uniform pair-wise correlations $\rho = \beta^2 / (1 + \beta^2)$. For these correlations not to be small, we must have $\beta^2\gsim 1$. Now, $Q = 1 + q$, where $q = N\beta^2$. For large $N$ we have $q\gg 1$, $Q\approx q$, and in this case we have a weighted regression over the intercept.

{}So, absent ``clustering", when $N$ is large, the factor model is only useful to the extent of defining the regression weights $z_i$ via the specific risks $\xi_i$. Thus, FCM does not affect the regression residuals: they are invariant under linear transformations of $\beta_{iA}$, (in matrix notations) $\beta\rightarrow\beta~U$, where $U_{AB}$ is a general nonsingular matrix.

{}What about ``clustering"? For a large number of alphas trading largely overlapping universes (e.g., top 2,500 most liquid U.S. stocks) there is no ``clustering" so long as they cannot be classified in a binary fashion as in industry classifications for stocks, and such a classification of alphas usually is not possible. Any risk factors then lack ``clustering" and are analogous to style factors or principal components.

\section{Deformed Sample Covariance Matrix}\label{sec.4}

{}Let us now discuss deforming -- or regularizing -- SCM such that it is nonsingular. One method often used in this regard is the so-called shrinkage \cite{LW}. It is often regarded as an ``alternative" to multifactor risk models. However, as was recently discussed in \cite{Shrunk}, shrunk SCM is also a factor model.

{}In fact, shrinkage is a special case of more general deformations, where instead of SCM $C_{ij}$ given by (\ref{SCM}), one uses
\begin{equation}\label{def}
 {\widetilde C}_{ij} = \Delta_{ij} + \sum_{s,s^\prime=1}^{M} X_{is}~{\widetilde \phi}_{ss^\prime}~X_{js^\prime}
\end{equation}
Here the matrix $\Delta_{ij}$ is assumed to be positive-definite and (relatively) stable out-of-sample. We must have ${\widetilde C}_{ii} = C_{ii}$, so this imposes $N$ conditions on $\Delta_{ii}$. A priori $\Delta_{ij}$ can be otherwise arbitrary. The matrix ${\widetilde \phi}_{ss^\prime}$ is some deformation of $\phi_{ss^\prime}$ in (\ref{SCM}).\footnote{\, In shrinkage, when translated into our language here, one simply takes ${\widetilde \phi}_{ss^\prime} = \left(1-\zeta\right) \phi_{ss^\prime}$ and (``shrinkage target") $\Delta_{ij} = \zeta~\Gamma_{ij}$, where the weight (``shrinkage constant") $0\leq \zeta\leq 1$, and $\Gamma_{ij}$ (usually chosen as a diagonal matrix or a $K$-factor model with low $K$) is such that $\Gamma_{ii}= C_{ii}$.\label{fn.shrunk}}

{}The issue with (\ref{def}) is that: i) in practice $\Delta_{ij}$ must have some relevance to the underlying returns whose covariance matrix we are attempting to model; and ii) a priori it is unclear what the deformed matrix ${\widetilde \phi}_{ss^\prime}$ should be. The available data is limited to the $N \times M$ matrix $X_{is}$, and the matrix $\phi_{ss^\prime}$, which is fixed. To go beyond this data, we invariably must introduce some additional input. As a 12th century Georgian poet Shota Rustaveli put it, ``What's in the jar is what will flow out."\footnote{\, This is ZK's own translation of an aphorism from a stanza in Rustaveli's sole known epic poem whose title is erroneously translated as ``The Knight in the Panther's Skin" (or similar). In ZK's humble opinion, not only is it the greatest masterpiece of the Georgian literature, but one of the greatest literary writings of all time. It consists of over 1,600 perfectly rhymed {\em shairi} or Rustavelian quatrains all containing $16 = 8 + 8$ syllables per line with a caesura between the 8th and 9th syllables. How a human brain can come up with such perfection is mind-boggling, especially considering that this poem tells an extremely complex story complete with dialogs, aphorisms, etc.}

{}However, not all is lost. The fact that we have large $N$ (and no ``clustering") simplifies things. In practice, the matrix $\Delta_{ij}$ cannot be arbitrary. It must be somehow -- be it directly or indirectly -- related to the returns whose covariance matrix we are after. The simplest choice is a diagonal matrix $\Delta_{ij} = D_i~\delta_{ij}$. More generally, we can take $\Delta_{ij}$ to be a $K$-factor model of the form (\ref{fac.mod}), $\Delta_{ij} = \Gamma_{ij}$, with a properly chosen diagonal $\Gamma_{ii}$. In fact, realistically, what else can $\Delta_{ij}$ be in practice? If we knew how to write down a non-factor-model covariance matrix that approximates SCM well and is out-of-sample stable, this paper would have been very different!

{}So, assuming $\Delta_{ij}$ is a $K$-factor model (with $K=0$ corresponding to a diagonal $\Delta_{ij}$) given by (\ref{fac.mod}), the deformed matrix ${\widetilde C}_{ij}$ is also a factor model. Indeed,
\begin{equation}\label{fac.mod.1}
 {\widetilde C}_{ij} = \xi_i^2~\delta_{ij} + \sum_{\alpha,\beta=1}^{K+M} {\widehat \Omega}_{i\alpha}~{\widehat \Phi}_{\alpha\beta}~{\widehat \Omega}_{j\beta}
\end{equation}
Here: the index $\alpha = (A, s)$ takes $K+M$ values; ${\widehat \Omega}_{iA} = \Omega_{iA}$; ${\widehat \Omega}_{is} = X_{is}$, $s = 1,\dots, M$; ${\widehat\Phi}_{AB} = \Phi_{AB}$, $A,B=1,\dots,K$; ${\widehat \Phi}_{ss^\prime} = {\widetilde\phi}_{ss^\prime}$, $s,s^\prime = 1,\dots, M$; and ${\widehat\Phi}_{As} \equiv 0$.

{}Now we are in good shape. Indeed, assuming large $N$ and no ``clustering", we know that optimization using ${\widetilde C}_{ij}$ (instead of $C_{ij}$) in (\ref{weights}) -- a factor model -- reduces to a weighted regression of the expected returns over the $N\times(K+M)$ factor loadings matrix ${\widehat \Omega}_{i\alpha}$, {\em irrespective} of FCM $\Phi_{AB}$ or the deformed matrix ${\widetilde\phi}_{ss^\prime}$, and the latter we do not even have a (constrained enough) guiding principle for computing. All we need is to somehow compute the regression weights $z_i=1/\xi_i^2$, i.e., the specific risks.

\subsection{What about Regression Weights?}

{}The specific risks follow from (\ref{fac.mod.1}). However, to compute them, we do need to know $\Phi_{AB}$ and ${\widetilde\phi}_{ss^\prime}$. Indeed, recalling that ${\widetilde C}_{ii} = C_{ii}$, we have\footnote{\, Nontrivial algorithms are required to ensure that all $\xi_i^2$ so computed are positive and consistent with FCM. Such algorithms and source code are given in \cite{KYa} (see below).}
\begin{equation}\label{spec.risk}
 \xi_i^2 = C_{ii} - \sum_{A=1}^K \Omega_{iA}~\Phi_{AB}~\Omega_{iB} - \sum_{s=1}^M X_{is}~{\widetilde\phi}_{ss^\prime}~X_{is^\prime}
\end{equation}
So, as far as the deformed matrix ${\widetilde\phi}_{ss^\prime}$ is concerned, a priori we have $M(M+1)/2$ parameters to play with and not much guidance to play the game. In fact, there is no magic bullet here. Simplicity is essentially the only beacon we can follow...

{}Since at the end we have a weighted regression, which in itself does not require knowing $\Phi_{AB}$ or ${\widetilde\phi}_{ss^\prime}$ provided we know the weights, we can simply take $\xi_i^2 = C_{ii}$. This might appear to contradict (\ref{spec.risk}), but it does not. This is because the residuals $\varepsilon_i$ are invariant under the rescalings of the weights $z_i\rightarrow \lambda~z_i$, where $\lambda > 0$. So, setting $\xi_i^2 = C_{ii}$ is equivalent to setting $\xi_i^2 = \zeta~ C_{ii}$, where $0<\zeta<1$, which simply puts $N$ conditions on $K(K+1)/2$ (from $\Phi_{AB}$) plus $M(M+1)/2$ (from ${\widetilde\phi}_{ss^\prime}$) a priori unknowns. This system may appear to be overconstrained for large enough $N$, but there always exists a ``solution":\footnote{\, This is shrinkage with a diagonal ``shrinkage target" (see footnote \ref{fn.shrunk}).} we can simply take $K=0$ and ${\widetilde\phi}_{ss^\prime} = \left(1-\zeta\right)\phi_{ss^\prime}$.

{}So, can we have weights other than the inverse sample variances? A priori the answer is yes. Here is a simple prescription. As above we can set ${\widetilde\phi}_{ss^\prime} = \left(1-\zeta\right)\phi_{ss^\prime}$, but take $\Gamma_{ij}$ to be a nontrivial factor model ($K > 0$). We must have $\Gamma_{ii} = \zeta~C_{ii}$. Generally, $\xi_i^2$ need not equal rescaled $C_{ii}$. There is a notable exception: if we take\footnote{\, As in \cite{LW}.} $\Gamma_{ij}$ to have a uniform correlation matrix. Let the correlation be $\rho$. Then we have $\Gamma_{ij} = \zeta\sigma_i\sigma_j\left[\left(1-\rho\right)\delta_{ij} + \rho u_i u_j\right]$, where $u_i\equiv 1$ is the unit $N$-vector. In this case we have $\xi_i^2 = \zeta\left(1-\rho\right)C_{ii}$. So, in this 1-factor model the weights are the same as the inverse sample variances, albeit the regression is over $M+1$ columns.\footnote{\, To wit, the $M$ columns in $X_{is}$, $s=1,\dots,M$, plus a single column equal $\sigma_i$. Usually, there is a high correlation between the latter and a linear combination of the former (see below). In fact, we will argue below that the factor proportional to $\sigma_i$ should be taken out altogether, i.e., removed from the factor loadings matrix ${\widehat \Omega}_{i\alpha}$, irrespective of how the latter is constructed (see Section \ref{sec.5}).} If we take a different factor model (even a 1-factor model with nonuniform correlations), generally $\xi_i^2$ do not equal rescaled $C_{ii}$. So, what should/can the risk factor(s) be?

\subsection{Candidates for Additional Risk Factors}

{}Since we are assuming no ``clustering", i.e., there is no binary classification we can construct for our returns,\footnote{\, Such a classification has a factor loadings matrix (or a subset of its columns) of the form $\Omega_{iA} = \omega_i~\delta_{G(i), A}$, where $G:\{1,\dots,N\}\rightarrow \{1,\dots,K\}$ maps our $N$ returns to $K$ ``clusters". Note, however, that the ``weights" $\omega_i$ (not to be confused with the portfolio weights $w_i$ in (\ref{weights})) can be arbitrary (including negative) and need not be proportional to the unit $N$-vector $u_i\equiv 1$.} a priori there are two evident choices for what the additional $K$ risk factors can be: i) principal components and ii) style factors (analogous to those in equity risk models). Below we will discuss a 3rd possibility.

\subsubsection{Principal Components}

{}The idea is to take the first $K < M$ principal components of SCM $C_{ij}$ as $\Omega_{iA}$. More precisely, there is another choice, to wit, to take $\Omega_{iA} = \sigma_i~V^{(A)}_i$, $A=1,\dots,K$, where $V^{(a)}_i$, $a=1,\dots,N$, are the principal components of the sample correlation matrix $\Psi_{ij}=C_{ij}/\sigma_i\sigma_j$. Typically, the difference between the two choices is not make-it-or-break-it, with the latter preferred (and usually producing better results) as it factors out the (skewed, quasi-log-normally distributed) volatility $\sigma_i$ and deals with the principal components of $\Psi_{ij}$, whose off diagonal elements take values in the interval $(-1,1)$ and have a tight distribution. So, we will adapt this approach here.

{}As above, ${\widetilde\phi}_{ss^\prime} = \left(1-\zeta\right)\phi_{ss^\prime}$, we take $\Phi_{AB} = \zeta~\lambda^{(A)}~\delta_{AB}$, so our deformed SCM\footnote{\, Recall that ${\widetilde C}_{ii} = C_{ii}$, and $C_{ij} = \sigma_i\sigma_j\sum_{a=1}^M \lambda^{(a)}~V_i^{(a)}~V_j^{(a)}$.}
\begin{equation}
 {\widetilde C}_{ij} = \xi_i^2~\delta_{ij} + \sigma_i\sigma_j\sum_{a=1}^K \lambda^{(a)}~V_i^{(a)}~V_j^{(a)} + \sigma_i\sigma_j\left(1 - \zeta\right)\sum_{a=K+1}^M \lambda^{(a)}~V_i^{(a)}~V_j^{(a)}
\end{equation}
Here: $\lambda^{(a)}$ are the eigenvalues corresponding to the principal components $V_i^{(a)}$ ($\lambda^{(1)} \geq \lambda^{(2)} \geq \dots\geq \lambda^{(M)}$, and $\lambda^{(a)} \equiv 0$ for $a > M$); and $\xi_i^2 = \zeta\sigma_i^2\sum_{a=K+1}^M \lambda^{(a)}~[V_i^{(a)}]^2$. In terms of the regression, this construction only affects the regression weights $z_i = 1/\xi_i^2$. Indeed, the regression over the first $M$ principal components $V^{(a)}_i$, $a=1,\dots,M$, is the same as the regression over $X_{is}$, $s = 1,\dots,M$, as these two matrices are related to each other via a linear transformation $V_i^{(a)} = \sum_{s=1}^M X_{is}~U_{sa}$, where $U_{sa}$ is a nonsingular $M\times M$ matrix. As to $\xi_i^2$, calculating it requires computing the first $K$ principal components.\footnote{\, Note that $\xi_i^2/\zeta\sigma_i^2 = 1 - \sum_{a=1}^K \lambda^{(a)}~[V_i^{(a)}]^2$; the $a>1$ terms are weighted by smaller eigenvalues.\label{fn.xi}} For sufficiently low $K\ll M$ we can use the power iterations method \cite{PowerIt} (see Subsection 4.1 for the algorithm and Appendix B for R source code in \cite{KYb}).\footnote{\, This costs ${\cal O}(n_{iter}MN)$ operations, where the number of iterations $n_{iter} \gg K$. As $K$ increases, at some point $n_{tot}\gsim M$ and it makes more sense to use the next method.} If $K\sim M$, then we can use a no-iterations method (see Subsection 4.2 for the algorithm and Appendix C for R source code in \cite{KYb}).\footnote{\, This costs ${\cal O}(M^2N)$ operations.} E.g., we can calculate the regression weights by computing the first few principal components (which are different from the inverse sample variances) and then run a weighted regression of $E_i$ over $X_{is}$. Note, however, that for $N\gg 1$ the 1st principal component typically has a large cross-sectional correlation with the intercept, i.e., $\theta = {1\over \sqrt{N}}\sum_{i=1}^N V^{(1)}_i$ is close to 1, so if we take $K=1$, $\xi_i^2$ are close to rescaled $C_{ii}$. Furthermore, higher principal component terms are subleading (see footnote \ref{fn.xi}).\footnote{\, I.e., in the 0th approximation $\xi_i^2$ based on principal components are still close to rescaled $C_{ii}$.}

\subsubsection{Style Factors}\label{sub.sub.style}

{}Style factors are based on measured or estimated properties of our returns. Even in the case of stocks, their number is at most of order 10. In the case of alphas the a priori possible style factors are logs of volatility, turnover, momentum\footnote{\, Assuming momentum is positive; otherwise, we can use momentum over volatility, so its distribution is not too skewed. If momentum equals realized return, then this is the Sharpe ratio.} and, possibly, capacity\footnote{\, However, capacity is difficult to implement and it is unclear if it adds value.} \cite{AlphaFM}. We will discuss the first three below.

{}There are two parts to the story here. First and foremost, if $M \gg 1$, then on general grounds it is clear that adding a few style factors to the regression cannot make a big difference provided that we keep the regression weights fixed. Second, the 3 style factors above turn out to be poor predictors for pair-wise correlations. For turnover this was argued based on empirical evidence in \cite{KLT}. A similar analysis for volatility yields the same conclusion, that log of volatility is not a good predictor for pair-wise correlations.\footnote{\, Following \cite{KLT}, we define $\nu_i = \ln(\sigma_i/\mu)$, where $\mu$ is such that $\nu_i$ has zero mean. We define three symmetric tensor combinations $x_{ij}=u_i u_j$, $y_{ij}=u_i\nu_j + u_j\nu_i$, and $z_{ij}=\nu_i\nu_j$ ($u_i\equiv1$ is the unit $N$-vector). We further define a composite index $\{a\}=\{(i,j)|i>j\}$, which takes $L=N(N-1)/2$ values, i.e., we pull the off-diagonal lower-triangular elements of a general symmetric matrix $G_{ij}$ into a vector $G_a$.  This way we can construct four $L$-vectors $\Psi_a$, $x_a$, $y_a$ and $z_a$. Now we can run a linear regression of $\Psi_a$ over $x_a$, $y_a$ and $z_a$. Note that $x_a \equiv 1$ is simply the intercept (the unit $L$-vector), so this is a regression of $\Psi_a$ over $y_a$ and $z_a$ with the intercept. The results based on the same data as in \cite{KLT} are summarized in Table \ref{table.corr.vol} and Figure 1 confirming our conclusion above.\label{fn.reg}} Momentum is defined as an average realized return over some period of time. The expected return is also defined as an average realized return over some -- possibly other -- period of time. Allocating capital into alphas inherently is a ``momentum" strategy: usually one does not bet against alphas that have performed well in the past.\footnote{\, This does not necessarily mean that ``hockey-stick" alphas (i.e., those that have performed well in the past but have ``flat-lined") are not used in constructing a portfolio of alphas.} Including momentum in the factor loadings matrix in the regression (partly) ``kills" alpha.\footnote{\, If we replace volatility by momentum defined as the realized return over the entire period of the data sample in \cite{KLT}, log of momentum too turns out a poor predictor for pair-wise correlations. Table \ref{table.corr.mom} and Figure 2 summarize the regression results.}

{}One place where the style factors can make a difference is in computing the regression weights. I.e., we do not include them in the regression, but take the weights to be $z_i = 1/\xi_i^2$, where the specific risks $\xi_i$ are for a factor model based on the style factors only. That is, we model the correlation matrix $\Psi_{ij}$ via a factor model based on all or some of the four factors, the intercept, log of volatility, log of turnover, and log of momentum. As mentioned above, the intercept by itself yields rescaled sample variances, but makes a difference in combination with other factors.

{}We can always simply take the inverse sample variances as the regression weights and not bother with computing the specific risks based on the style factors. Without a detailed analysis using real-life alphas it is unclear if style factor based regression weights add value. Regardless, we need not include the style factors in the regression.

\subsubsection{How to Compute Specific Risk?}

{}Here we discuss the simplest -- albeit neither the only nor necessarily the best -- way to compute the specific risks. As above, instead of modeling SCM $C_{ij}$ via a factor model, it is convenient to model the sample correlation matrix $\Psi_{ij}$. Let the corresponding factor model be
\begin{equation}\label{fac.mod.2}
 {\widetilde \Gamma}_{ij} = {\widetilde \xi}_i^2~\delta_{ij} + \sum_{A,B=1}^K {\widetilde \Omega}_{iA}~\Phi_{AB}~{\widetilde \Omega}_{jB}
\end{equation}
where ${\widetilde\Gamma}_{ij} = \Gamma_{ij}/\sigma_i\sigma_j$, ${\widetilde\xi}_i = \xi_i/\sigma_i$, and ${\widetilde\Omega}_{iA} = \Omega_{iA}/\sigma_i$. First, without loss of generality we can assume that the columns of ${\widetilde \Omega}_{iA}$ are linearly independent. Second, we can assume that they form an orthonormal basis, i.e., the matrix $H_{AB} = \sum_{i=1}^N {\widetilde\Omega}_{iA}~{\widetilde\Omega}_{iB}$ is the $K\times K$ identity matrix: $H_{AB} = \delta_{AB}$. Indeed, we can always ensure orthonormality via the transformation (in matrix notations) ${\widetilde \Omega} \rightarrow {\widetilde \Omega}~({\widetilde H}^T)^{-1}$, where ${\widetilde H}$ is the Cholesky decomposition of $H$, so ${\widetilde H}~{\widetilde H}^T = H$. Furthermore, we have ${\widetilde\Gamma}_{ii} = \Psi_{ii} \equiv 1$.

{}With orthonormal ${\widetilde \Omega}_{iA}$, FCM $\Phi_{AB}$ is simply a projection of the sample correlation matrix onto the $K$-dimensional hyperplane defined by the columns of ${\widetilde\Omega}_{iA}$ in the $N$-dimensional space:\footnote{\, See, e.g., \cite{Het} or \cite{KYa} for a more detailed discussion.}
\begin{equation}\label{FCM}
 \Phi_{AB} = \sum_{i,j=1}^N {\widetilde\Omega}_{iA}~\Psi_{ij}~{\widetilde\Omega}_{jB}
\end{equation}
The specific risks then follow from (\ref{fac.mod.2}):
\begin{equation}\label{xi.twiddle}
 {\widetilde\xi}_i^2 = {\widetilde \Gamma}_{ii} - \sum_{A,B=1}^K {\widetilde \Omega}_{iA}~\Phi_{AB}~{\widetilde \Omega}_{iB} = 1 - \sum_{A,B=1}^K {\widetilde \Omega}_{iA}~\Phi_{AB}~{\widetilde \Omega}_{iB}
\end{equation}
However, there is a caveat in this approach. For a generic matrix ${\widetilde\Omega}_{iA}$ there is no guarantee that the so-defined ${\widetilde\xi}_i^2$ are positive, which they should be. This imposes nontrivial restrictions on ${\widetilde\Omega}_{iA}$. As an illustration, let us discuss the $K=1$ case.

{}For $K=1$ things simplify. Let us denote the sole column of ${\widetilde\Omega}_{iA}$ via $\beta_i$. Then $\sum_{i=1}^N\beta_i^2 = 1$ and
\begin{equation}\label{sp.risk}
 {\widetilde\xi}_i^2 = 1 - \kappa~\beta_i^2
\end{equation}
where $\kappa = \sum_{i,j=1}^N \beta_i~\Psi_{ij}~\beta_j \leq \lambda^{(1)}$, and $\lambda^{(1)}$ is the largest eigenvalue of $\Psi_{ij}$. So, a sufficient condition for having all ${\widetilde\xi}_i^2 > 0 $ is that all $\beta_i^2 < 1/\lambda^{(1)}$. We can replace this condition with a stronger one that avoids computing the largest eigenvalue: a sufficient condition is that all $\beta_i^2 \leq 1/{\lambda_*}$, where ${\lambda_*} = {1\over N} \sum_{i,j=1}^N\Psi_{ij}$. Note that $\beta_i \equiv 1/\sqrt{N}$, which corresponds to the intercept as the sole risk factor, satisfies this condition. Violations of this condition usually occur for $\beta_i$ with a skewed distribution, e.g., if $\beta_i \propto \sigma_i$; however, for, e.g., $\beta_i \propto \ln(\sigma_i)$ such violations are either absent altogether or rarer and can be dealt with by ``reigning" in the few violating elements. See the R function {\tt{\small qrm.fr()}} in Appendix A of \cite{KYa} for such an algorithm, which is built for a general $K$-factor model (not just $K=1$).

{}When we have multiple factors, however, things get trickier. Even if ${\widetilde \xi}_i^2 > 0$ in all $K$ 1-factor models based on the individual columns of a multifactor ${\widetilde\Omega}_{iA}$, in the $K$-factor model we can still have some ${\widetilde \xi}_i^2 < 0$. See Section 4 for the algorithm and Appendix B for R source code in \cite{KYa} for circumventing this issue (even for $K=1$). However, let us note one issue associated with computing the specific risks even if they all come out to be positive. Computing ${\widetilde \xi}_i^2$ via (\ref{xi.twiddle}) involves FCM $\Phi_{AB}$, which in turn involves the sample correlation matrix $\Psi_{ij}$ via (\ref{FCM}). While $\Phi_{AB}$ is expected to be more out-of-stable than $\Psi_{ij}$ as we have $K\ll N$, using FCM in (\ref{xi.twiddle}) still adds some noise to the regression weights. This is to be contrasted with using the inverse sample variances (i.e., ${\widetilde\xi}_i^2 \equiv 1$), which are relatively stable out-of-sample, as the regression weights. This remark equally applies to all $K$ values.

\subsection{Can We Increase the Number of Factors?}

{}So, while we can try to play with the regression weights, it is not all that clear that this would make it or break it, especially that we are still limited to the $M$ risk factors, which are equivalent to the first $M$ principal components -- albeit we never have to compute the principal components in the first instance. The question is, can we increase the number of columns in the factor loadings matrix in the regression as this presumably would cover more directions in the risk space and improve the out-of-sample performance. Prosaically, the answer is that we need more information, i.e., additional input, to achieve this. Here is one approach \cite{AlphaFM}.

{}The idea here is that, assuming all alphas have essentially overlapping trading universes, we can treat exposure to each underlying tradable -- for the sake of definiteness let us assume we are dealing with the U.S. equities as the underlying tradables -- as a risk factor. This makes sense, but the question is what should the factor loadings matrix ${\widetilde \Omega}_{iA}$ be? In this case $A$ simply labels stocks in the trading universe, which is, say, top 2,500 most liquid tickers, so $K$ is large, much larger than the typical value of $M$, which for a (generous)\footnote{\, Many alphas can be more ephemeral than that.} 1-year lookback is only about 250.

{}Historical stock position data for each alpha must be available to us if we are to backtest them. Let this position data be $P_{iAs}$, which is the dollar holding of the alpha labeled by $i$ in the stock labeled by $A$ at time labeled by $t_s$, normalized such that $\sum_A |P_{iAs}| = 1$ for each given pair $i,s$. We can try to construct ${\widetilde \Omega}_{iA}$ from $P_{iAs}$ by getting rid of the time series index $s$. The most obvious choice ${\widetilde \Omega}_{iA} = {1\over{M+1}}\sum_{s=1}^{M+1} P_{iAs}$ does not work as the sign of $P_{iAs}$ flips over time frequently assuming alphas have reasonably short holding periods. We need an unsigned quantity to define ${\widetilde \Omega}_{iA}$. We can use\footnote{\, The overall normalization factor is immaterial and included for aesthetic reasons.}
\begin{equation}\label{modulus.1}
 {\widetilde \Omega}_{iA} = {1\over{M+1}}\sum_{s=1}^{M+1} \left|P_{iAs}\right|
\end{equation}
This is simply average relative exposure of the $i$-th alpha to the stock labeled by $A$.

{}One potential ``shortcoming" of this definition is that, if the position bounds -- call them $B_{iA}$ -- are imposed at the level of individual alphas, for some, mostly less liquid, stocks $\left|P_{iAs}\right|$ could be saturating these bounds. On general risk management grounds, these bounds can very well be uniform across all alphas. E.g., one may wish to cap the positions as the smaller of: i) a small percentile of the total dollar investment -- this is a diversification bound; and ii) a (generally, different) small percentile of ADDV (average daily dollar volume) -- this is a liquidity bound (in case positions need to be liquidated). If the bounds are saturated most of the time, this can effectively reduce the number of independent risk factors, and the definition of ${\widetilde \Omega}_{iA}$ may have to be modified for such stocks (see \cite{AlphaFM}). However, if the bounds are imposed at the level of the combined alpha, then this is a non-issue.

{}Assuming $N\gg K$, even with the larger number of risk factors (\ref{modulus.1}), our optimization reduces to a weighted regression. We can simply choose the weights as the inverse sample variances. Alternatively, we can attempt to compute specific risks. For this we need FCM $\Phi_{AB}$. With appropriately normalized ${\widetilde \Omega}_{iA}$, FCM $\Phi_{AB}$ is simply the covariance matrix for the stocks. In the zeroth approximation we can set it to $\Phi_{AB}\approx \sigma_A^2~\delta_{AB}$, where $\sigma_A^2$ are sample variances for stocks. Alternatively, we can either use commercial risk models or construct them organically, as, e.g., in \cite{Het} and \cite{KYa}. Also see Section \ref{sec.6} hereof.

\section{A Refinement}\label{sec.5}

{}So, to summarize, our optimization (\ref{weights}) reduces to a regression of the normalized returns ${\widetilde E}_i = E_i/\xi_i$ over the factor loadings matrix ${\widetilde \Omega}_{iA} = \Omega_{iA}/\xi_i$. For simplicity, let us take $\xi_i = \sigma_i$. Further, let us take $\Omega_{iA}$ to be the $M$ demeaned returns $X_{is}$, $s = 1,\dots, M$, i.e., we are not using any additional style or other risk factors. Then the sample correlation matrix is given by (we identify the index $A$ with the index $s$)
\begin{equation}\label{Psi}
 \Psi_{ij} = \sum_{s,s^\prime=1}^M {\widetilde\Omega}_{is}~\phi_{ss^\prime}~{\widetilde\Omega}_{js^\prime}
\end{equation}
It is evident that the columns of ${\widetilde\Omega}_{is}$ are nothing but some linear combinations of the first $M$ principal components $V^{(a)}_i$, $a=1,\dots,M$, of $\Psi_{ij}$. For large $N$ the first principal component $V^{(1)}_i$ is close to the appropriately normalized unit $N$-vector: $V_i^{(1)} \approx 1/\sqrt{N}$. Recall from (\ref{w.reg}) (see the discussion right thereafter) that the weights $w_i \approx \eta~{\widetilde\varepsilon}_i/\sigma_i$, where ${\widetilde \varepsilon}_i$ are the residuals of the regression of ${\widetilde E}_i = E_i/\sigma_i$ over ${\widetilde\Omega}_{is}$ with unit weights, or, equivalently, over the principal components $V^{(a)}_i$. This implies that $\sum_{i=1}^N V^{(1)}_i~{\widetilde\varepsilon}_i = 0$ and, therefore, many weights $w_i$ are negative (assuming the expected returns $E_i$ are all positive). I.e., the vanilla optimization forces us to take bets against many positive expected returns because it hedges against all alphas simultaneously losing money. This is overkill and literally ``kills" alpha: if alphas are not highly correlated on average, most alphas losing money all at once is possible but highly unlikely. Assuming tolerance for drawdowns, we can relax this hedge.

\subsection{A 1-factor Example}

{}To further illustrate this point, let us consider a simple example. Let us hypothetically assume that the true correlation matrix has uniform off-diagonal elements: $\Psi_{ij} = \left(1-\rho\right)\delta_{ij} + \rho u_i u_j$, where $u_i\equiv 1$ is the unit $N$-vector. Then the weights are given by (note that this is exact):
\begin{equation}\label{w.1F}
 w_i = {\eta\over\xi_i}\left[{\widetilde E}_i - {\rho\over{1 + \left(N-1\right)\rho}}\sum_{i=1}^N {\widetilde E}_i\right]
\end{equation}
where ${\widetilde E}_i = E_i/\xi_i$, and $\xi_i = \sqrt{1-\rho}~\sigma_i$. So, if $\rho > 0$ and $N\gg 1/\rho$, then the expression in the square brackets on the r.h.s. of (\ref{w.1F}) is approximately equal cross-sectionally demeaned ${\widetilde E}_i$ and many weights will be negative.\footnote{\, The distribution of $\sigma_i$ is skewed and roughly log-normal, and so is that of $E_i$, so the distribution of ${\widetilde E}_i$ is not very skewed and is roughly normal; see \cite{4K}.} How can we fix this?

\subsection{Removing the ``Overall" Mode}

{}In the case of equity portfolios with a large number of tickers we also have the same behavior, that the first principal component of the sample correlation matrix is close to the rescaled intercept. This is known as the ``market" mode and describes the overall, in-sync movement of all stocks, i.e., of the broad market as a whole.\footnote{\, See, e.g., \cite{CFM}, which reviews applications of random matrix theory to modeling a sample correlation matrix for equities, and references therein.} However, in the case of equity portfolios the issue of the ``market" mode is subdued. Thus, for dollar-neutral portfolios roughly half of the weights are negative by the dollar-neutrality constraint. In long-only portfolios one typically does not optimize the raw expected returns directly but against a broad benchmark. In this regard, alpha portfolio optimization is analogous to long-only portfolio optimization for equities. So, here too we could optimize against a benchmark alpha portfolio (as opposed to (\ref{weights})). E.g., we can take
\begin{equation}
 w_i^{benchmark} = \eta~{E_i\over\sigma_i^2}
\end{equation}
which corresponds to taking the diagonal part of SCM $C_{ij}$. There are other choices.

{}Alternatively, we can simply remove the ``overall" mode (i.e., the analog of the ``market" mode for equity portfolios). One way is to take the $M$ principal components and simply remove the first principal component, i.e., to run the regression over the factor loadings matrix ${\widetilde \Omega}_{ia}^\prime = V^{(a)}_i$, $a=2,\dots,M$. However, this would require computing the principal components. There is a simpler way. We take ${\widetilde \Omega}_{is}$ in (\ref{Psi}) and demean its columns. Let the resulting matrix be ${\widetilde\Omega}_{is}^*$. Only $M-1$ of its columns are linearly independent. So, we can regress over ${\widetilde\Omega}_{is}^*$, $s=1,\dots,M-1$. This removes the ``overall" mode and while some weights may still turn out to be negative, their number will be relatively limited (not roughly half of the weights). The backtested Sharpe ratio will go down, and the portfolio return will go up.

\subsection{Summary of Regression Procedure}\label{sub.reg.algo}

{}For clarity, let us put all the pieces together into a step-by-step summary:\footnote{\, Here we deliberately use slightly different notations than above.}

{}$\bullet$ 1) Start with a time series of alpha returns\footnote{\, As before, $i=1,\dots,N$ labels the alphas; $s=1,\dots,M+1$ labels the times $t_s$.}  $R_{is}$, $i=1,\dots,N$, $s=1,\dots,M+1$.

{}$\bullet$ 2) Calculate the serially demeaned returns $X_{is} = R_{is} - {1\over {M+1}}\sum_{s=1}^{M+1} R_{is}$.

{}$\bullet$ 3) Calculate sample variances\footnote{\, Their normalization is immaterial in what follows.} $\sigma_i^2 = C_{ii} = {1\over M}\sum_{s=1}^{M+1} X^2_{is}$.

{}$\bullet$ 4) Calculate the normalized demeaned returns $Y_{is} = X_{is} /\sigma_i$.

{}$\bullet$ 5) Keep only the first $M$ columns in $Y_{is}$: $s=1,\dots, M$.

{}$\bullet$ 6) Cross-sectionally demean\footnote{\, This step removes the ``overall" mode and can be skipped if so desired. Then we would also skip the next step 7) below.} $Y_{is}$: $\Lambda_{is} = Y_{is} - {1\over N}\sum_{j=1}^N Y_{js}$.

{}$\bullet$ 7) Keep only the first $M-1$ columns in $\Lambda_{is}$: $s=1,\dots,M-1$.

{}$\bullet$ 8) Take the alpha expected returns $E_i$ and normalize them: ${\widetilde E}_i = E_i/\sigma_i$.

{}$\bullet$ 9) Calculate the residuals ${\widetilde \varepsilon}_i$ of the unit-weighted regression\footnote{\, Without the intercept.} of ${\widetilde E}_i$ over $\Lambda_{is}$.

{}$\bullet$ 10) Set the alpha portfolio weights to $w_i = \eta~ {\widetilde\varepsilon}_i/\sigma_i$.

{}$\bullet$ 11) Set the normalization coefficient $\eta$ such that $\sum_{i=1}^N \left|w_i\right| = 1$.

{}Source code in R for the above procedure is given in Appendix \ref{app.A}. If we wish to use the underlying tradables as the risk factors as in (\ref{modulus.1}), then we simply replace $\Lambda_{is}$ in step 9) above by the corresponding factor loadings matrix (see Appendix \ref{app.A}).\footnote{\, Which can be, {\em e.g.}, (\ref{modulus.1}) or a union thereof with $Y_{is}$ defined in step 5) above (with any linearly dependent columns removed).} Also, in steps 4)-11) above instead of using sample variances $\sigma_i^2$, we can use specific variances $\xi_i^2$ computed based on principal components or style factors (see above).

\subsection{What about Computational Cost?}

{}It is evident than none of the steps above cost more than ${\cal O}(MN)$ operations except perhaps for step 9), the regression. The regression residuals are given by
\begin{equation}
 {\widetilde\varepsilon}_i = {\widetilde E}_i - \sum_{j=1}^N \sum_{s,s^\prime=1}^{M-1} \Lambda_{is}~\Upsilon_{ss^\prime}^{-1}~\Lambda_{js^\prime}
\end{equation}
where $\Upsilon_{ss^\prime} = \sum_{i=1}^N \Lambda_{is}~\Lambda_{is^\prime}$. Calculating this $(M-1)\times (M-1)$ matrix costs ${\cal O}(M^2 N)$ operations. Straightforwardly inverting it costs ${\cal O}(M^3)$ operations.\footnote{\, The fact that $\Upsilon_{ss^\prime}$ has a factor form does not help as $N \gg M$, in fact, in practice $N\gg M^2$.} The rest (sums over $s, s^\prime, j$, etc.) costs ${\cal O}(M^2 N)$ operations, and therefore so does the regression.

\subsection{Is This Related to Principal Components?}\label{sub.pc}

{}The answer is yes. As we discussed above, the sample correlation matrix $\Psi_{ij} = \sum_{s,s^\prime = 1}^M Y_{is}~\phi_{ss^\prime}~Y_{js^\prime}$ (where $\phi_{ss^\prime} = \left(\delta_{ss^\prime} + u_s u_{s^\prime}\right)/M$; $u_s\equiv 1$ is the unit $M$-vector), so the $M$ columns of $Y_{is}$ are just linear combinations of the first $M$ principal components $V^{(a)}_i$, $a=1,\dots,M$, of $\Psi_{ij}$. So, we could use the principal components instead of $Y_{is}$ in the above steps and get the same result\footnote{\, We can keep step 6) above; alternatively, we can simply drop the first principal component, which will give a slightly different set of $w_i$.} for the weights $w_i$. However, computing the first $M$ principal components costs additional ${\cal O}(M^2 N)$ operations.

\section{Conclusions}\label{sec.6}

{}As we discussed above, in the absence of ``clustering", when the number of alphas is large, optimization (via maximizing the Sharpe ratio) reduces to a (weighted) regression irrespective of whether we start from a constructed factor model, or deform the sample covariance matrix. This is because such deformations themselves are nothing but factor models. We also argued that in most cases the factor loadings, over which the expected returns are regressed, are given by the (properly demeaned, normalized and trimmed) time series matrix of historical returns based on which the (singular) sample covariance matrix is computed. The regression weights, which can be recast as the normalizations of the expected returns, the factor loadings matrix and the alpha weights, can be taken as inverse sample variances or, alternatively, specific variances in some factor model. However, computation of these specific variances via (\ref{xi.twiddle}) requires computing the factor covariance matrix via (\ref{FCM}) (thereby adding noise to the regression weights) not needed to compute sample variances.

{}There is a notable exception to this, to wit, if we use the underlying tradables (stocks) themselves as risk factors via (\ref{modulus.1}). In this case the factor covariance matrix $\Phi_{AB}$ need not be computed via a sample covariance matrix of linear combinations of the alpha returns. Instead, it can be taken to be the covariance matrix for stocks. The latter need not be computed as a sample covariance matrix of stock returns, which would be out-of-sample unstable or, even worse, singular. Instead, we can use a constructed covariance matrix for stocks, e.g., via a factor model \cite{AlphaFM}. A priori we could use commercially available risk models (albeit they are not necessarily expected to be out-of-sample stable), or build them organically via, e.g., heterotic risk models \cite{Het} or heterotic CAPM \cite{KYa}. In fact, in the zeroth approximation we can set $\Phi_{AB}\approx \sigma^2_A~\delta_{AB}$, where $\sigma_A$ are sample (historical) stock volatilities or implied volatilities from options.\footnote{\, Albeit not all stocks in the trading universe may be optionable and have impled volatilities readily available. Also, it is unclear whether implied volatilities add value \cite{Het}.} And once we nail down $\Phi_{AB}$, we can compute the specific variances via (\ref{xi.twiddle}). However, in reality there is a caveat here. The caveat is that if we identify $\Phi_{AB}$ with the stock covariance matrix, then the factor loadings matrix is given by (\ref{modulus.1}) only up to an overall normalization constant which is a priori unknown. So we can treat it as a free parameter and consider a 1-parameter family of specific variances. The value of this parameter then can be fixed by optimizing for realized performance with the caveat that it need not be out-of-sample stable and may have to be recomputed frequently based on short lookbacks. This provides a well-defined prescription for computing specific variances. Alternatively, we can use sample variances, which are relatively stable out-of-sample and simple to compute. Yet another alternative is to use specific variances computed based on principal components (see \cite{KYb} and above) or style factors. The latter case generally requires using more sophisticated methods such as those discussed in \cite{KYa}.

{}A nice thing about our optimization reducing to a regression, which is computationally cheap, is that it can be readily modified to incorporate bounds on the alpha weights $w_i$. Indeed, since $w_i$ are proportional to the regression residuals, we can simply use the bounded regression discussed in\cite{Bounded}. Similarly, we can incorporate trading costs via the method discussed in \cite{TCosts}.

{}Finally, let us mention that if we use the algorithms of \cite{KYb} for building statistical risk models, which include fixing the number of risk factors (i.e., principal components) $K$ together with the specific risks, we are going to end up with a regression over $K < M$ principal components with the regression weights equal the inverse specific variances. In the case of stocks this works better than regressing over $M$ principal components with the regression weights ad hoc set equal the inverse specific variances for a $K$-factor model with $K < M$ \cite{KYb}. Same may or may not hold for, e.g., $N \sim 100,000$ real-life alphas.

\appendix

\section{R Code for Alpha Weights}\label{app.A}

{}In this appendix we give the R source code for calculating the alpha weights based on a regression. The code below is essentially self-explanatory and straightforward as it simply follows the algorithm and formulas in Subsection \ref{sub.reg.algo}. It consists of a single function {\tt{\small calc.opt.weights(e.r, ret, y = 0, s = 0, rm.overall = T)}}; {\tt{\small e.r}} is an $N$-vector of expected returns we wish to optimize; $N$ is the number of the underlying returns (e.g., alphas); {\tt{\small ret}} is an $N\times (M+1)$ matrix of returns; $M+1$ is the number of data points in the time series (e.g., days); {\tt{\small y}} is an $N\times K$ factor loadings matrix ${\widetilde \Omega}_{iA}$, $A=1,\dots,K$, pre-computed, e.g., via (\ref{modulus.1}); otherwise, if the default {\tt{\small y = 0}} is used, the code computes the factor loadings matrix $Y_{is}$ ($s=1,\dots,M-1$ or $s=1,\dots,M$ depending on whether {\tt{\small rm.overall = T}} or  {\tt{\small rm.overall = F}} -- see below) based on the time series {\tt{\small ret}} via the algorithm of Subsection \ref{sub.reg.algo}; {\tt{\small s}} is an $N$-vector of specific risks ${\widetilde\xi}_i$ pre-computed, e.g., via (\ref{xi.twiddle}) or (\ref{sp.risk}); otherwise, if the default {\tt{\small s = 0}} is used, the code computes {\tt{\small s}} as the square root of the sample variances; {\tt{\small rm.overall}}, if {\tt{\small TRUE}} (default), implies that the ``overall" mode is taken out; otherwise, it is kept (see Subsection \ref{sub.reg.algo}). The output is an $N$-vector $w_i$ of the optimized alpha weights normalized such that $\sum_{i=1}^N\left|w_i\right| = 1$. The code can be easily modified, e.g., to combine (via {\tt{\small cbind()}}) the pre-computed factor-loadings matrix ${\widetilde \Omega}_{iA}$ as in (\ref{modulus.1}) with the factor loadings matrix $Y_{is}$ it already computes based on the time series {\tt{\small ret}}.\\
\\
{\tt{\small
\noindent calc.opt.weights <- function(e.r, ret, y = 0, s = 0, rm.overall = T)\\
\{\\
\indent if(length(s) == 1)\\
\indent \indent s <- apply(ret, 1, sd)\\
\\
\indent if(length(y) == 1)\\
\indent \{\\
\indent \indent x <- ret - rowMeans(ret)\\
\indent \indent y <- x / s\\
\indent \indent y <- y[, -ncol(x)]\\
\indent \}\\
\\
\indent if(rm.overall)\\
\indent \{\\
\indent \indent y <- t(t(y) - colMeans(y))\\
\indent \indent y <- y[, -ncol(y)]\\
\indent \}\\
\\
\indent e.r <- matrix(e.r / s, length(e.r), 1)\\
\indent w <- t(y) \%*\% e.r\\
\indent w <- solve(t(y) \%*\% y) \%*\% w\\
\indent w <- e.r - y \%*\% w\\
\indent w <- w / s\\
\indent w <- w / sum(abs(w))\\
\indent return(as.vector(w))\\
\}
}}

\section{DISCLAIMERS}\label{app.C}

{}Wherever the context so requires, the masculine gender includes the feminine and/or neuter, and the singular form includes the plural and {\em vice versa}. The author of this paper (``Author") and his affiliates including without limitation Quantigic$^\circledR$ Solutions LLC (``Author's Affiliates" or ``his Affiliates") make no implied or express warranties or any other representations whatsoever, including without limitation implied warranties of merchantability and fitness for a particular purpose, in connection with or with regard to the content of this paper including without limitation any code or algorithms contained herein (``Content").

{}The reader may use the Content solely at his/her/its own risk and the reader shall have no claims whatsoever against the Author or his Affiliates and the Author and his Affiliates shall have no liability whatsoever to the reader or any third party whatsoever for any loss, expense, opportunity cost, damages or any other adverse effects whatsoever relating to or arising from the use of the Content by the reader including without any limitation whatsoever: any direct, indirect, incidental, special, consequential or any other damages incurred by the reader, however caused and under any theory of liability; any loss of profit (whether incurred directly or indirectly), any loss of goodwill or reputation, any loss of data suffered, cost of procurement of substitute goods or services, or any other tangible or intangible loss; any reliance placed by the reader on the completeness, accuracy or existence of the Content or any other effect of using the Content; and any and all other adversities or negative effects the reader might encounter in using the Content irrespective of whether the Author or his Affiliates is or are or should have been aware of such adversities or negative effects.

{}The R code included in Appendix \ref{app.A} hereof is part of the copyrighted R code of Quantigic$^\circledR$ Solutions LLC and is provided herein with the express permission of Quantigic$^\circledR$ Solutions LLC. The copyright owner retains all rights, title and interest in and to its copyrighted source code included in Appendix \ref{app.A} hereof and any and all copyrights therefor.

\newpage

\begin{table}[ht]
\noindent
\caption{Summary for the cross-sectional regression of $\Psi_a$ over $y_a$ and $z_a$ with the intercept, where $y_a$ and $z_a$ are based on log of volatility. See Subsection \ref{sub.sub.style} for details. Also see Figure 1.}
\begin{tabular}{l l l l l} 
\\
\hline\hline 
 & Estimate & Standard error & t-statistic & Overall \\[0.5ex] 
\hline 
Intercept & 0.1588 &  0.0016 &   97.28  & \\
$y_a$ & 0.0331 &   0.0029 &   11.37  &   \\ 	
$z_a$ & 0.1354 &  0.0106 &   12.82    & \\		
Mult./Adj. R-squared & & & & 0.0540~/~0.0536  \\
F-statistic	& & & & 144.0 \\ [1ex] 
\hline 
\end{tabular}
\label{table.corr.vol} 
\end{table}

\begin{table}[ht]
\noindent
\caption{Summary for the cross-sectional regression of $\Psi_a$ over $y_a$ and $z_a$ with the intercept, where $y_a$ and $z_a$ are based on log of momentum. See Subsection \ref{sub.sub.style} for details. Also see Figure 2.}
\begin{tabular}{l l l l l} 
\\
\hline\hline 
 & Estimate & Standard error & t-statistic & Overall \\[0.5ex] 
\hline 
Intercept & 0.1587 &   0.0017 &  95.74    & \\
$y_a$ & 0.0158 &   0.0033 &   4.74 &   \\ 	
$z_a$ & 0.1389 &   0.0137 &  10.14      & \\		
Mult./Adj. R-squared & & & & 0.0238~/~0.0234   \\
F-statistic	& & & & 61.58 \\ [1ex] 
\hline 
\end{tabular}
\label{table.corr.mom} 
\end{table}

\newpage
\begin{figure}[ht]
\centerline{\epsfxsize 4.truein \epsfysize 4.truein\epsfbox{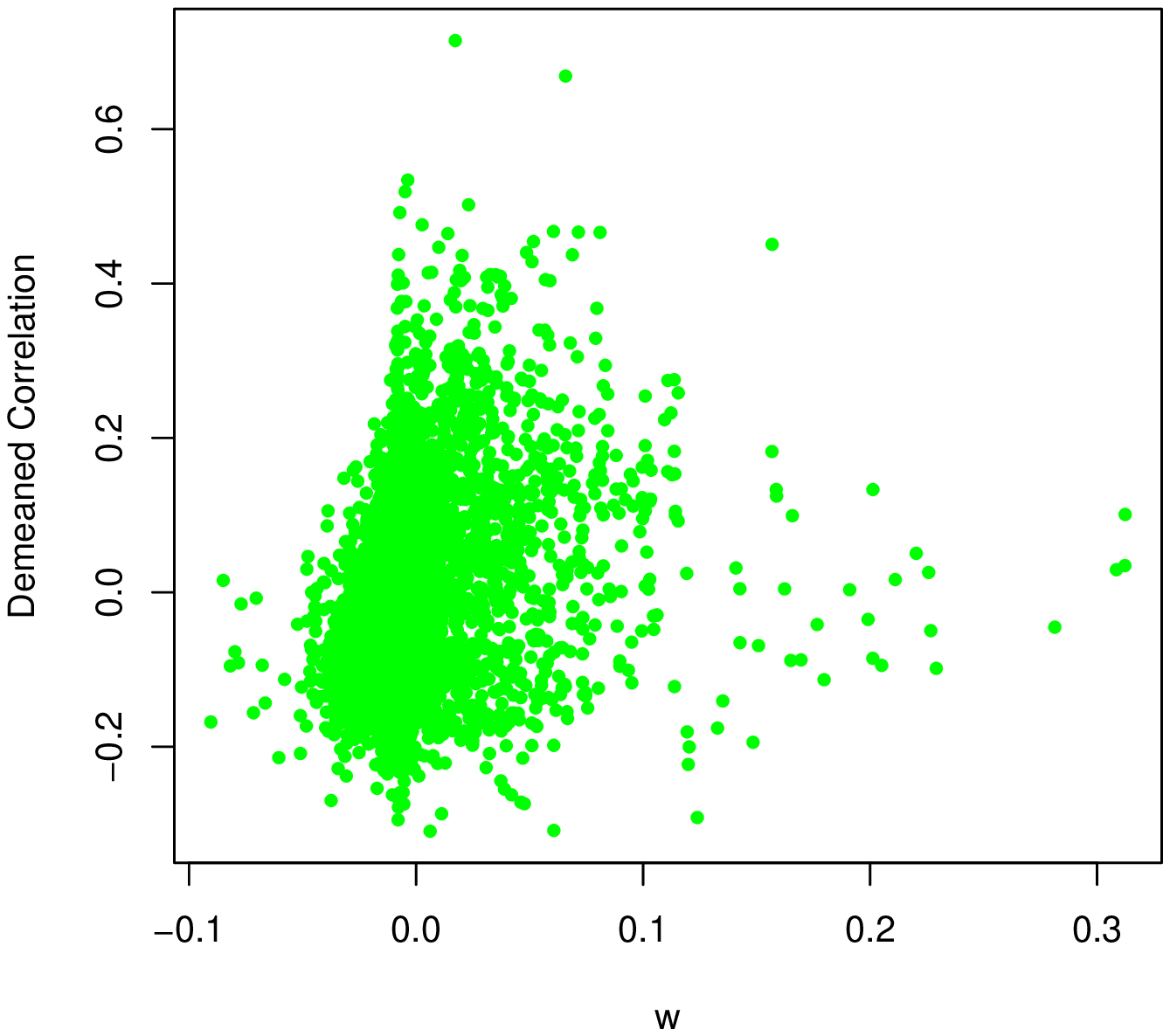}}
\noindent{\small {Figure 1. Horizontal axis: $w_a = 0.0331~y_a + 0.1354~z_a$; vertical axis: $\Psi_a - \mbox{Mean}(\Psi_a)$. See
Table \ref{table.corr.vol} and Subsection \ref{sub.sub.style}. The numeric coefficients in $w_a$ are the regression coefficients in Table \ref{table.corr.vol}.}}
\end{figure}

\newpage
\begin{figure}[ht]
\centerline{\epsfxsize 4.truein \epsfysize 4.truein\epsfbox{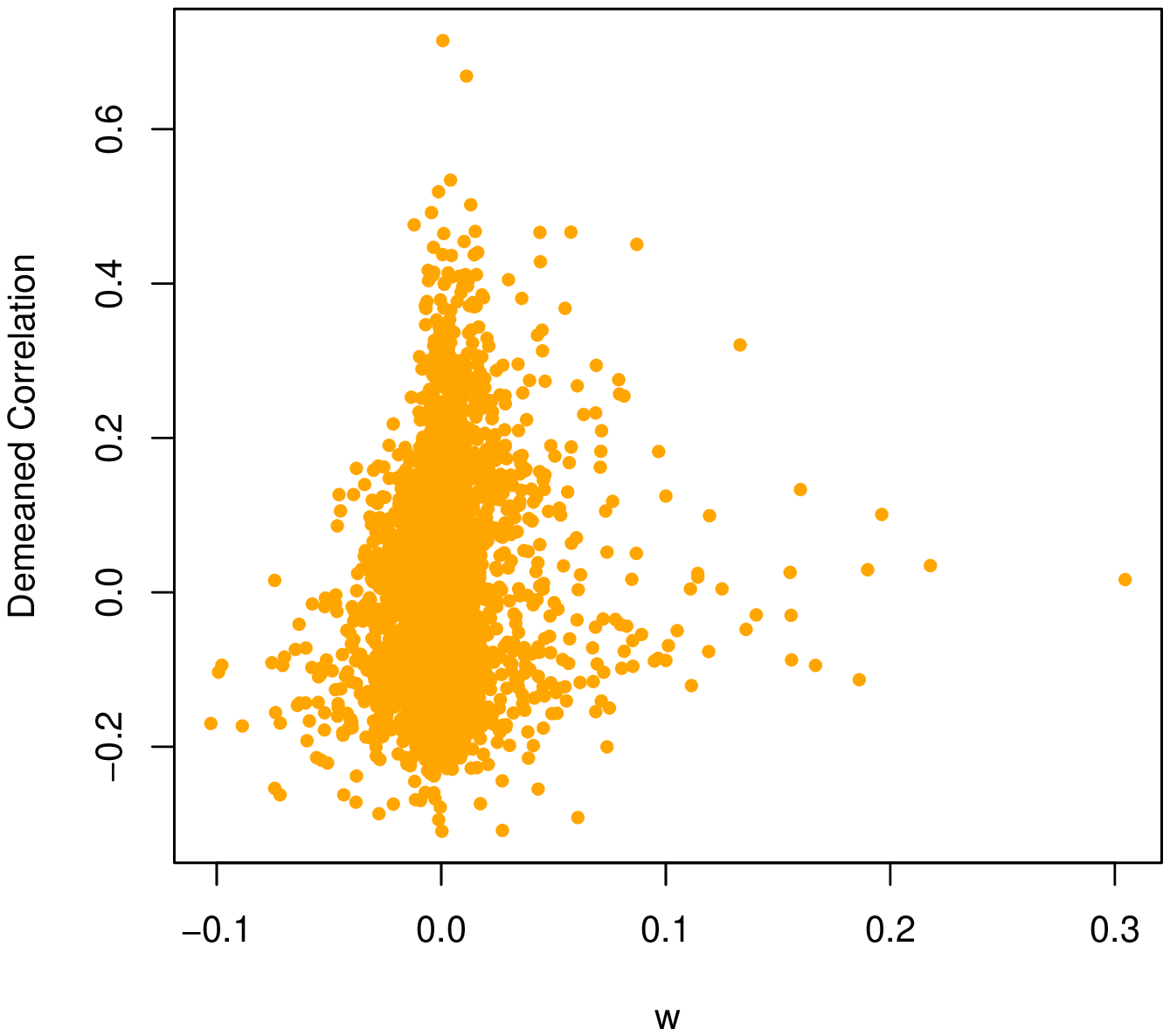}}
\noindent{\small {Figure 2. Horizontal axis: $w_a = 0.0158~y_a + 0.1389~z_a$; vertical axis: $\Psi_a - \mbox{Mean}(\Psi_a)$. See
Table \ref{table.corr.mom} and Subsection \ref{sub.sub.style}. The numeric coefficients in $w_a$ are the regression coefficients in Table \ref{table.corr.mom}.}}
\end{figure}


\begin{thebibliography}{99}

\makeatletter
\def\@biblabel#1{}
\makeatother

\bibitem[Bouchaud and Potters, 2011]{CFM} Bouchaud, J.-P. and Potters, M.
``Financial applications of random matrix theory: a short review." In: Akemann, G., Baik, J. and Di Francesco, P. (eds.) The Oxford Handbook of Random Matrix Theory. Oxford, United Kingdom: Oxford University Press, 2011.

\bibitem[Grinold and Kahn, 2000]{GK} Grinold, R.C. and Kahn, R.N.
``Active Portfolio Management." New York, NY: McGraw-Hill, 2000.

\bibitem[Kakushadze, 2014]{AlphaFM} Kakushadze, Z.
``Factor Models for Alpha Streams."
The Journal of Investment Strategies, 4(1) (2014), pp. 83-109.\\
Available online: http://ssrn.com/abstract=2449927.

\bibitem[Kakushadze, 2015a]{TCosts} Kakushadze, Z.
``Combining Alpha Streams with Costs." The Journal of Risk, 17(3) (2015a), pp. 57-78.
Available online: http://ssrn.com/abstract=2438687.

\bibitem[Kakushadze, 2015b]{Bounded} Kakushadze, Z.
``Combining Alphas via Bounded Regression." Risks, 3(4) (2015b), pp. 474-490. Available online: http://ssrn.com/abstract=2550335.

\bibitem[Kakushadze, 2015c]{Het} Kakushadze, Z.
``Heterotic Risk Models."
Wilmott Magazine, 2015(80) (2015c), pp. 40-55. Available online: http://ssrn.com/abstract=2600798.

\bibitem[Kakushadze, 2015d]{KLT}
Kakushadze, Z. ``101 Formulaic Alphas."
Wilmott Magazine (forthcoming). Available online: http://ssrn.com/abstract=2701346 (2015d).

\bibitem[Kakushadze, 2016]{Shrunk} Kakushadze, Z.
``Shrinkage = Factor Model."
Journal of Asset Management, (17)(2) (2016), pp. 69-72. Available online: http://ssrn.com/abstract=2685720.

\bibitem[Kakushadze and Tulchinsky, 2016]{4K} Kakushadze, Z. and Tulchinsky, I.
``Performance v. Turnover: A Story by 4,000 Alphas."
The Journal of Investment Strategies, 5(2) (2016), pp. 75-89. Available online: http://ssrn.com/abstract=2657603.

\bibitem[Kakushadze and Yu, 2016a]{KYa} Kakushadze, Z. and Yu, W.
``Multifactor Risk Models and Heterotic CAPM."
The Journal of Investment Strategies, 5(4) (2016a) (forthcoming). Available online: http://ssrn.com/abstract=2722093.

\bibitem[Kakushadze and Yu, 2016b]{KYb} Kakushadze, Z. and Yu, W.
``Statistical Risk Models." Working Paper. Available online: http://ssrn.com/abstract=2732453 (2016b).

\bibitem[Ledoit and Wolf, 2004]{LW} Ledoit, O. and Wolf, M.
``Honey, I Shrunk the Sample Covariance Matrix."
The Journal of Portfolio Management, 30(4) (2004), pp. 110-119.

\bibitem[Markowitz, 1952]{Markowitz} Markowitz, H.
``Portfolio selection."
The Journal of Finance, 7(1) (1952), pp. 77-91.

\bibitem[Mises and Pollaczek-Geiringer, 1929]{PowerIt} Mises, R.V. and Pollaczek-Geiringer. H.
``Praktische Verfahren der Gleichungsaufl\"osung."
ZAMM -- Zeitschrift f\"ur Angewandte Mathematik und Mechanik, 9(2) (1929), pp. 152-164.

\bibitem[Sharpe, 1994]{Sharpe94} Sharpe, W.F.
``The Sharpe Ratio."
The Journal of Portfolio Management, 21(1) (1994), pp. 49-58.

\end{thebibliography}
\end{document}